# Comment on 'A Dynamic ID-based Remote User Authentication Scheme

Amit K Awasthi

**Abstract** — *Since 1981, when Lamport introduced the remote user authentication scheme using table, a plenty of schemes had been proposed with table and without table using. Recently Das, Saxena and Gulati have proposed A dynamic ID-based remote user authentication scheme. They claimed that their scheme is secure against ID-theft, and can resist the reply attacks, forgery attacks, insider attacks and so on.*
*In this paper we show that Das et al.'s scheme is completely insecure and using of this scheme is like an open server access without password. 1*

## I. INTRODUCTION

REMOTE user authentication schemes allow a valid user to login to the remote server and to access the services provided. This authentication process runs over the insecure channels. Lamport's table based authentication scheme was enhanced in 2000, by Hwang and Li [7], to remote user authentication scheme using smart cards. Afterwards many schemes have been proposed to make secure the authentication over insecure channels.

---
[1] A. K. Awasthi is with the Department of Applied Science, Hindustan College of Science and Technology, Farah, Mathura, UP, INDIA. (E-mail: awasthi_hcst@ yahoo.com)

Recently Das et al. proposed a dynamic ID-based authentication scheme. In this scheme they introduce the concept of dynamic ID to overcome the problem of partial information leakage in static ID based schemes. This also avoids the risk of ID-theft.

In this paper, we shall point out that the Das et al. scheme is completely insecure. We shall show that in 'Login Phase' of this scheme the send login request is password independent. User may type any random password instead of a real one. Scheme does not prevent him from login.

In section II we review the Das et al's scheme. Section III consists of the comments on the scheme. Finally in section IV a brief conclusion is given.

## II. REVIEW OF THE DAS ET AL. SCHEME

In this section, we briefly review Das et al. scheme [5]. This scheme is composed of the registration phase, Login phase, authentication phase and the password change phase. The notations used are as follows:

| | |
|---|---|
| $U$ | the user |
| $PW$ | the password of user $U$ |
| $S$ | the remote server |
| $h(.)$ | a one way hash function |
| $\oplus$ | bitwise XOR operation |

$A \Rightarrow B: M$  $A$ sends $M$ to $B$ over secure channel
$A \rightarrow B : M$  $A$ sends $M$ to $B$ over insecure channel

Different phases work as follows:

## A. Registration Phase

A user $U_i$ wants to register to the remote system $S$.

1. $U_i$ submits $PW_i$ to $S$
2. $S$ computes $N_i = h(PW_i) \oplus h(x)$, where x is secret of the remote system.
3. $S$ Personalizes the smartcard with the parameters $[h(.),\ N_i,\quad y\ ]$, where $y$ is a remote server's secret number stored in each registered user's smartcard.
4. $S \Rightarrow U_i$ : $PW_i$ and smartcard.

## B. Login Phase

The user wants to login, inserts its smartcard to the terminal and keys his password $PW_i$. The smartcard perform the following steps:

1. Computes $CID_i = h(PW_i) \oplus h(N_i \oplus y \oplus T)$, where $T$ is the current date and time.
2. Computes $B_i = h(CID_i \oplus h(PW_i))$
3. Computes $C_i = h(T \oplus N_i \oplus\ B_i \oplus y)$
4. $U_i \to S$: $CID$, $N_i$, $C_i$, $T$

## C. Authentication Phase

Upon receiving the login request $(CID, N_i, C_i, T)$ at time $T^*$, $S$ verifies as :

1. Verify the validity of the time interval $T - T^*$
2. Computes $h(PW_i) = CID_i \oplus h(N_i \oplus y \oplus T)$
3. Computes $B_i = h(CID_i \oplus h(PW_i))$
4. Checks that $C_i = h(T \oplus N_i \oplus\ B_i \oplus y)$ holds to accept the login request.

## D. Password Change Phase

When user wants to change the password he inserts smartcard in to the device, keys thw password $PW_i$ and request to change the password to new one $PW_{New}$. Smartcard computes $N_i^* = N_i \oplus h(PW_i) \oplus\ h(PW_{New})$ and replaces the $N_i$ with new $N_i^*$. Password gets changed.

## III. Comment on Das et al. scheme

The Login and authentication phase is completely insecure, because whole process of authentication is independent of the password. Attack may work as –

## Login Phase

The user wants to login, inserts its smartcard to the terminal and keys a random password $P$ instead of his real password $PW_i$. The smartcard perform the following steps:

1. Computes $CID_i = h(P) \oplus h(N_i \oplus y \oplus T)$, where $T$ is the current date and time.

2. Computes $B_i = h(CID_i \oplus h(P))$
3. Computes $C_i = h(T \oplus N_i \oplus\ B_i \oplus y)$
4. $U_i \to S$: $CID$, $N_i$, $C_i$, $T$

## A. Authentication Phase

Upon receiving the login request $(CID, N_i, C_i, T)$ at time $T^*$, $S$ verifies as :

1. Verify the validity of the time interval $T - T^*$
2. Computes $h(P) = CID_i \oplus h(N_i \oplus y \oplus T)$
3. Computes $B_i = h(CID_i \oplus h(P))$
4. Checks that $C_i = h(T \oplus N_i \oplus\ B_i \oplus y)$ holds to accept the login request.

$C_i$ will hold true, this shows that this scheme is equivalent to no password scheme, because with any random password user may access the server.

Suppose an intruder theft the smartcard for a short duration and makes a duplicate of it. Now he has no need to crack the password because he may insert any random password. Server will authenticate the intruder as a valid user as discussed above.

## IV. Conclusion

In this paper we show that the Das et al.'s scheme is insecure and works like an open channel. No password is required to authenticate the user. This scheme does not fulfill the basic need of authentication schemes.
.

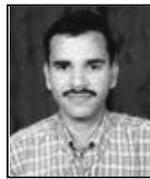

**Amit K Awasthi** received his M. Sc. Degree in 1999 from Bareilly College, (M. J. P. Rohilkhand University,) Bareilly. He is currently a lecturer in Department of Applied Science, Hindustan College of Science and Technology, Farah, Mathura, INDIA. He is member of Indian Mathematical Society, Group for Cryptographic Research, Cryptography Research Society of India and Computer Society of India. His current research interests include data security and cryptography.

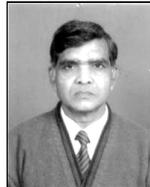

**Sunder Lal** is currently a professor and Head of Department of Mathematics, IBS Khandari, Dr. B. R. A. University, Agra, INDIA. He is member of Indian Mathematical Society, Group for Cryptographic Research, and Cryptography Research Society of India. His current research interests include cryptography, number theory and applied algebra.